\documentclass[prd,onecolumn]{revtex4}
\usepackage{dcolumn}
\usepackage{multirow}
\usepackage{graphicx}
\usepackage{amssymb}
\usepackage{bm}
\usepackage{hyperref}
\usepackage{epstopdf}
\usepackage{color}
\usepackage{mathrsfs}
\usepackage{amsmath,amssymb,amsthm}
\begin{document}
\title{Model-independent traversable wormholes from baryon acoustic oscillations}
\author{Deng Wang}
\email{cstar@nao.cas.cn}
\affiliation{Instituto de F\'{i}sica Corpuscular (CSIC-Universitat de Val\`{e}ncia), E-46980 Paterna, Spain \\
National Astronomical Observatories, Chinese Academy of Sciences, Beijing, 100012, China}
\begin{abstract}
In this paper, we investigate the model-independent traversable wormholes from baryon acoustic oscillations. Firstly, we place the statistical constraints on the average dark energy equation of state $\omega_{av}$ by only using BAO data. Subsequently, two specific wormhole solutions are obtained, i.e, the cases of the constant redshift function and a special choice for the shape function. For the first case, we analyze the traversabilities of the wormhole configuration, and for the second case, we find that one can construct theoretically a traversable wormhole with infinitesimal amounts of average null energy condition violating phantom fluid. Furthermore, we perform the stability analysis for the first case, and find that the stable equilibrium configurations may increase for increasing values of the throat radius of the wormhole in the cases of a positive and a negative surface energy density. It is worth noting that the obtained wormhole solutions are static and spherically symmetrical metric, and that we assume $\omega_{av}$ to be a constant between different redshifts when placing constraints, hence, these wormhole solutions can be interpreted as stable and static phantom wormholes configurations at some certain redshift which lies in the range [0.32, 2.34].

\end{abstract}
\maketitle

\section{Introduction}
New precise observations can reveal the origin of the accelerating universe: specifically, to determine if cosmic acceleration is caused by deviations from general relativity (GR) on the large scales or by a new form of energy (i.e., so-called dark energy). The theory that characterizes the deviation from GR is dubbed as modified gravity \cite{CANTATA:2021ktz}, which is highly degenerate with dark energy models from a theoretical viewpoint. It is possible to break the degeneracy between scenarios of acceleration that require dark energy components and those that require modifications to GR by independently explore both the cosmic expansion history and the structure growth rate. By using the cosmological observations at both the background and perturbation levels to constrain these theories, one can compress the theory space and rule out some models.  
There are four basic observational techniques accepted as the most powerful toward achieving that goal \cite{1}, namely, Type Ia supernovae (SNe Ia), weak lensing, galaxy clusters and Baryon acoustic oscillations (BAO). BAO that propagate in the pre-recombination universe imprint a characteristic scale in the clustering of matter components. The location of the peak provides a `` universal ruler '', which can be measured in the power spectrum of cosmic microwave background (CMB) anisotropies and in the maps of large scale structure (LSS) at low redshifts, with a constant comoving length $r_s$ ($150$ Mpc) to probe nearly all of the cosmic history. The sharpening of BAO precision with gradually increasing redshift, the difference between absolute and relative measurements and the independent systematic uncertainties make the BAO observations substantially effective tools to test various kinds of dark energy scenarios. Generally speaking, in spectroscopic surveys, one can directly obtain useful information of the Hubble parameter $H(z)$ by the BAO measurements in the line of sight, apart from the constraints from the clustering perpendicular to the line of sight on the comoving angular diameter distance $D_C(z)\propto \int^z_0cH^{-1}(z)dz$ in a spatially flat background geometry.

In this paper, we would like to focus on the first case and the two-point correlation function has a peak at $\Delta z=r_sH(z)/c$. Hence, the BAO observations can determine the combination $r_sH(z)$ for different redshifts. According to the recent paper \cite{2}, on the one hand, the resulting determination of $r_sH(z)$ is independent of the assumed cosmological scenario, since both packaging accumulated data in terms of just one redshift for each sample and determining the position of the peak by comparing the matter correlation functions with simulations have a small and negligible dependence on the fiducial cosmological scenario. On the other hand, the value of $r_s$ does depend on the assumed cosmological scenario. Furthermore, combing with recent observations, one can easily find that, even though this model dependence is not very large, it is already larger than the uncertainty obtained by several BAO measurements. Subsequently, it is substantially natural that one would like to know how to obtain the model independent information from the above mentioned physical quantities. J. Evslin has pointed out that one can probe the cosmic expansion history by ratios of BAO measurements \cite{2}. To be more precise, if knowing the location of the peak at different redshifts $z_1$ and $z_2$, one can obtain $r_sH(z_1)/c$ and $r_sH(z_2)/c$. Although each depends on the fiducial scenario individually, the ratio does not.

The baryon oscillations spectroscopic survey (BOSS) of SDSS-III has two main objectives \cite{3}: (i) to measure the BAO distance scale with $1\%$ precision from a survey of 1.5 million luminous galaxies at $z=0.2\sim0.7$; (ii) to make the first BAO measurement at $z>2$ taking 3-dimensional structure in the Ly$\alpha$ forest absorption towards a dense grid of 160,000 high redshift quasars. In this paper, we would like to utilize the final results from BOSS which give the $H(z)$ measurements for samples of galaxies in two redshift groups, i.e., the closer sample LOWZ and the farther sample CMASS. The results are well compatible with the base cosmology scenario (i.e., $\Lambda$CDM). In the meanwhile, we will also utilize the BOSS measurements of the BAO in both the auto-correlation of masses and cross-correlation of mass densities, which determine the Hubble parameter $H(z)$ at an effective redshift $z=2.34$. The related results are all listed in Table. \ref{tab1}. Through some detailed analysis, J. Evslin finds that, taking only the BOSS line of sight Ly$\alpha$ forest and CMASS measurements, there exists already more than $3\sigma$ tension with the base cosmology scenario which indicates that either (i) The BOSS Ly$\alpha$ forest measurement of $H(z)$ was too low due to a systematic error or statistical fluctuation or else (ii) the dark energy equation of state falls rapidly at high redshifts.
\begin{table}[h!]
\begin{tabular}{ccccccc}
\hline
                           &effective redshift $z$  & Measured $H(z)r_s/c$ & Ref.     \\
\hline
LOWZ             &$z=0.32$        &$0.038\pm0.0021$                     &  \cite{a1}        \\
CMASS                  &$z=0.57$       &$0.0485\pm0.0013$               &   \cite{a2}        \\
Ly$\alpha$                    &$z=2.34$       &$0.109\pm0.002$          &   \cite{a3}         \\
\hline
\end{tabular}
\caption{ The measurements of the position of the peak of the spatial two-point correlation $\Delta z=r_sH(z)/c$.}
\label{tab1}
\end{table}

In the previous works \cite{4,5,6}, we have studied the geometrical dark energy traversable wormholes supported by modern astrophysical observations and demonstrated that the exotic spacetime configurations traversable wormholes may actually exist in the universe. However, the previously obtained results depend strongly on concrete cosmological models. Therefore, one question comes into being naturally: whether do the model independent traversable wormholes exist in the real universe, namely, whether can one explore the existence of wormholes directly starting from astrophysical data?

To answer this question, in the present paper, we are dedicated to explore the model independent dark energy traversable wormholes supported by the current BAO observations, which are entirely different from our previous works \cite{4,5,6}. We would like to construct the average dark energy equation of state starting directly from BAO data, and investigate the corresponding properties and characteristics of the wormhole spacetime configurations. It is worth noting that the wormhole spacetime configurations in modified gravity theories have been deeply explored in recent several years, and the corresponding progresses can be found in Ref.\cite{Sharif:2018sgg,Sarkar:2020aaz,Fayyaz:2020jzh,Mustafa:2020qjo,Fayyaz:2020xey,FarasatShamir:2020csw,FarasatShamir:2021xzq,Godani:2022nkm,Bhattacharya:2023imf,Agrawal:2022atn,Jawad:2018axf,Arouko:2020gbi,Paul:2021lvb,Godani:2022tyw,Eid:2020hvf,Ghosh:2021msy,Godani:2019kgy,Wang:2016uun}

This paper is organized in the following manner. In Section 2, we will make a review on the average dark energy equation of state and exhibit our further analysis. In Section 3, two specific traversable wormhole solutions are presented and the related properties are also studied. In Section 4, we explore about the stability of the wormhole solutions. In the final section, the discussions and conclusions are presented (we use the units $8\pi G=c=1$).
\section{Review on the average dark energy equation of state}
Following \cite{2}, we would like to make an assumption that the universe is homogeneous and isotropic, and is full of a perfect fluid with pressure $p$ and density $\rho$ obeying an equation of state $\omega=p/\rho$. For the convenience of calculations in the following context, we will set the spatial curvature to be zero. Therefore, the first Friedmann equation is expressed as follows
\begin{equation}
H^2(z)=\frac{\rho(z)}{3}=\frac{\rho_M(z)+\rho_{DE}(z)}{3}, \label{1}
\end{equation}
where $\rho(z)$, $\rho_M(z)$, and $\rho_{DE}(z)$ are the total, non-relativistic matter, and dark energy to the density at redshift $z$, respectively. Furthermore, the evolution of the non-relativistic matter and the dark energy can be, respectively, expressed as
\begin{equation}
\rho_M(z)=\rho_0(0)(1+z)^3, \label{2}
\end{equation}
\begin{equation}
\rho_{DE}(z)=\rho_{DE}(0)e^{3\int^z_0\frac{1+\omega(z')}{1+z'}dz'}. \label{3}
\end{equation}
Specially, the dark energy density between the redshifts $z_1$ and $z_2$ can evolve in the following manner
\begin{equation}
\rho_{DE}(z_2)=\rho_{DE}(z_1)e^{3\int^{z_2}_{z_1}\frac{1+\omega(z')}{1+z'}dz'}. \label{4}
\end{equation}
Then, using Eq. (\ref{2}), the ratio of two measurements of the Hubble parameter $H(z)$ at different redshifts is written as
\begin{equation}
[\frac{H(z_1)}{H(z_2)}]^2=\frac{1+\frac{\rho_{DE}(z_1)}{\rho_M(z_1)}}{(\frac{1+z_2}{1+z_1})^3+\frac{\rho_{DE}(z_1)}{\rho_M(z_1)}e^{3\int^{z_2}_{z_1}\frac{1+\omega(z')}{1+z'}dz'}}. \label{5}
\end{equation}
It is worth noting that this equation gives a relationship between two quantities: $\rho_{DE}(z_1)/\rho_M(z_1)$ and $e^{3\int^{z_2}_{z_1}\frac{1+\omega(z')}{1+z'}dz'}$. The first one provides information about redshift $z_1$, and the second one about the evolution between $z_1$ and $z_2$. Since we are very interested in the evolutionary behavior of dark energy, we take into account the average equation of state
\begin{equation}
\omega_{av}(z_1,z_2)=-1+\frac{\int^{z_2}_{z_1}\frac{1+\omega(z')}{1+z'}dz'}{\ln(\frac{1+z_2}{1+z_1})}. \label{6}
\end{equation}
Subsequently, based on the concern in \cite{2}, one can obtain the general dark energy density as
\begin{equation}
\rho_{DE}(z_2)=(\frac{1+z_2}{1+z_1})^{3(1+\omega_{av}(z_1,z_2))}\rho_{DE}(z_1). \label{7}
\end{equation}
Substituting Eq. (\ref{7}) into Eq. (\ref{5}), one can have
\begin{equation}
[\frac{H(z_1)}{H(z_2)}]^2=\frac{(\frac{1+z_1}{1+z_2})^3(1+\frac{\rho_{DE}(z_1)}{\rho_M(z_1)})}{1+\frac{\rho_{DE}(z_1)}{\rho_M(z_1)}(\frac{1+z_2}{1+z_1})^{3\omega_{av}(z_1,z_2)}}. \label{8}
\end{equation}
One can see that this equation just contains two parameters, namely, $\rho_{DE}(z_1)/\rho_M(z_1)$ and $\omega_{av}(z_1,z_2)$. Therefore, it is convenient
to constrain these two parameters from a ratio of the Hubble parameter $H(z)$ based on the BAO observations. To perform a chi square statistical analysis, the theoretical ratio of the Hubble parameter at different redshifts is defined as
\begin{equation}
H_{th}=[\frac{H(z_1)}{H(z_2)}]^2, \label{9}
\end{equation}
and the corresponding $\chi^2$ can be denoted as
\begin{equation}
\chi^2={(\frac{H_{obs}-H_{th}}{\sigma})}^2, \label{10}
\end{equation}
where $H_{obs}$ and $\sigma$ denote the derived ratio of the Hubble parameters from BAO data and 1 statistical error, respectively.

\begin{figure}
\centering
\includegraphics[scale=0.3]{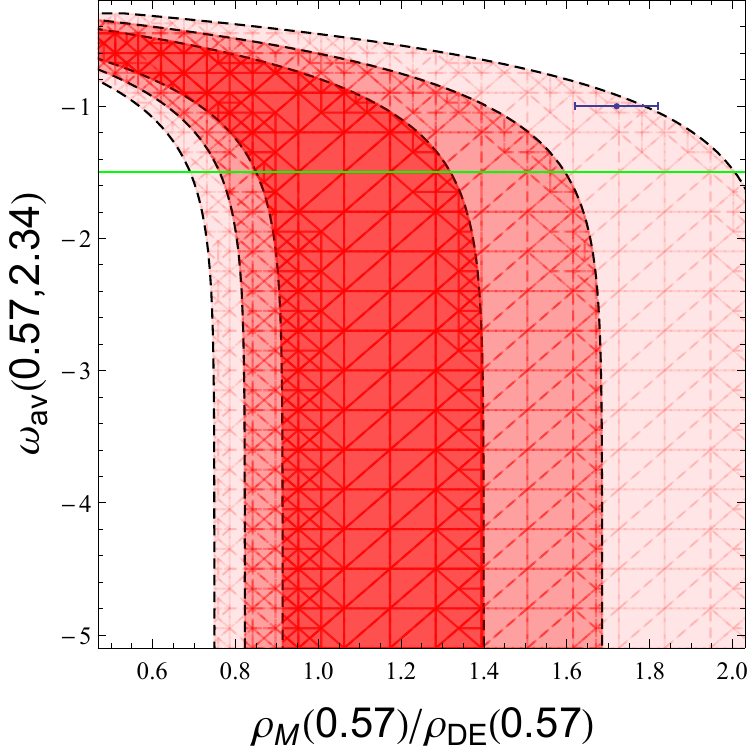}
\includegraphics[scale=0.3]{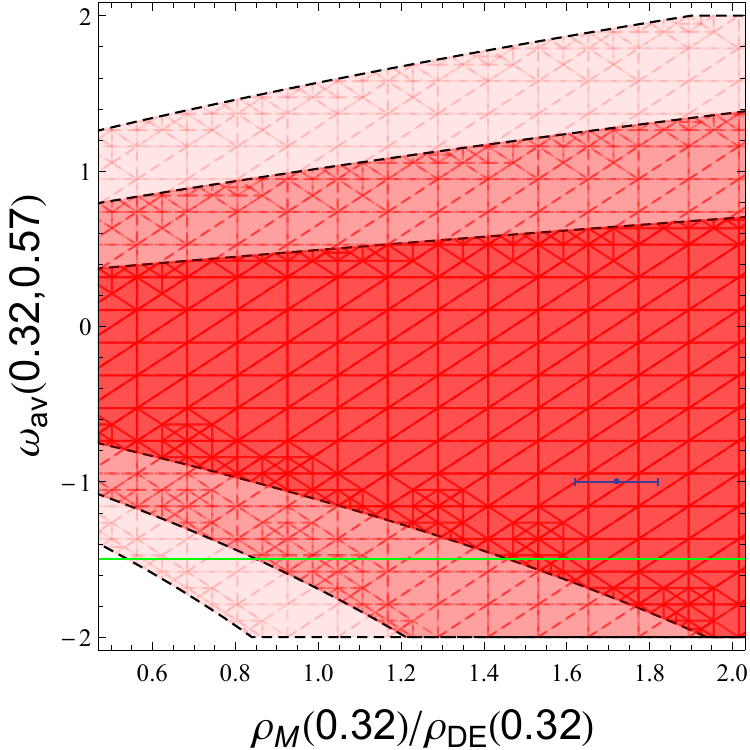}
\includegraphics[scale=0.3]{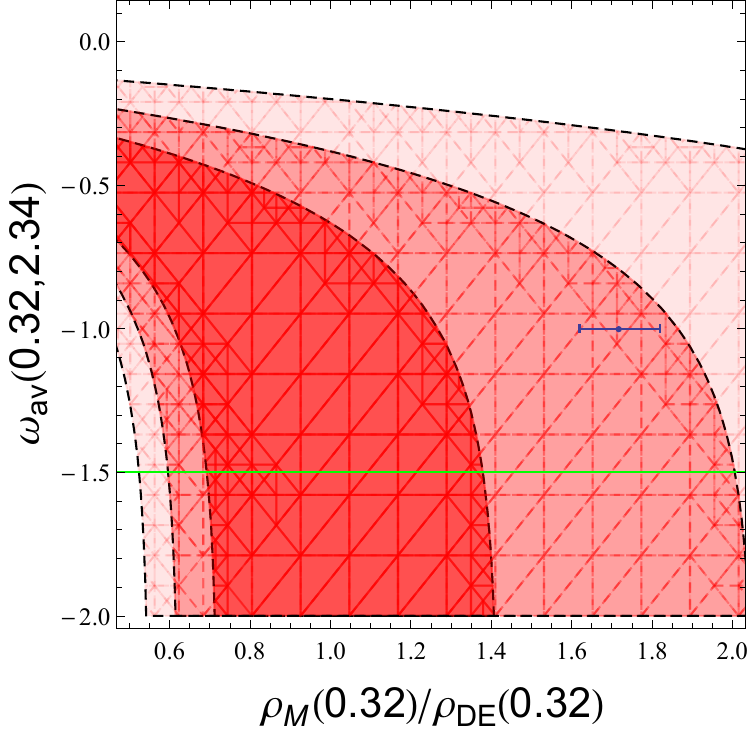}
\caption{The $1\sigma$, $2\sigma$ and $3\sigma$ confidence intervals for the parameter pairs (${\rho_M(0.57)}/{\rho_{DE}(0.57)},\omega_{av}(0.57,2.34)$), (${\rho_M(0.32)}/{\rho_{DE}(0.32)},\omega_{av}(0.32,0.57)$) and (${\rho_M(0.32)}/{\rho_{DE}(0.32)},\omega_{av}(0.32,2.34)$) are shown from left to right, respectively. The blue dots with 1$\sigma$ errorbar correspond to the recent Planck's result. The horizontal green (solid) lines correspond to the dark energy equation of state $\omega_{av}(0.57,2.34)=-1.5$.}\label{f1}
\end{figure}

According to Eq. (14) in \cite{2} and Planck's constraint result $\Omega_M=0.308\pm0.012$ \cite{P}, where $\Omega_M$ denotes present-day matter density parameter, the standard cosmological model with Planck's estimate of $\Omega_M$ yields
\begin{equation}
\frac{\rho_M(0.57)}{\rho_{DE}(0.57)}=1.72\pm0.10. \label{11}
\end{equation}
Subsequently, we can also have three ratios of the Hubble scales through some simple calculations
\begin{eqnarray}
  [\frac{H(0.57)}{H(2.34)}]^2&=&0.198\pm0.013, \\\nonumber
  [\frac{H(0.32)}{H(0.57)}]^2&=&0.64\pm0.08, \\\nonumber
  [\frac{H(0.32)}{H(2.34)}]^2&=&0.127\pm0.014 \label{12},
\end{eqnarray}
which are the same as those in \cite{2}. Furthermore, the constraint results are exhibited in Figs. \ref{f1}. It is easily to be seen that our results are different from those in \cite{2}, namely, in the left panel, our result is consistent with Planck's result at 3$\sigma$ level but the latter at over 3$\sigma$ level; in the medium panel, our result is consistent with Planck's result at 1$\sigma$ level but the latter at 2 $\sigma$ level; in the right panel, our result is consistent with Planck's result at 2$\sigma$ level but the latter at 1 $\sigma$ level. We think that the discrepancies can be attributed to the different choices of the flat prior for the parameter pair in the above-mentioned three cases. Note that all the way we choose the flat prior $[0, 10]$ for the ratio  $\rho_M(z_1)/\rho_{DE}(z_1)$ and $[-5, 2]$ for the average dark energy equation of state $\omega_{av}(z_1,z_2)$.

In the above, we have implemented the constraints on the average dark energy equation of state and the density ratio of matter and dark energy in three different cases. From Fig. \ref{f1}, one can easily find that the null energy condition (NEC, i.e., $T_{\mu\nu}k^{\mu}k^{\nu}>0$, where $T_{\mu\nu}$ is the stress-energy tensor and $k^{\mu}$ any future directed null vector.) is violated. More specifically, the NEC in this situation can be translated into $\omega_{av}(z_1,z_2)<-1$. Therefore, we can directly extract information about the average dark energy equation of state from the current BAO observations to probe the existences of traversable wormholes spacetime configurations. It is noteworthy to stress once again that we would like to investigate the model independent traversable wormholes, which are independent of any cosmic model and start directly from BAO data. In the next section, we think that it is substantially constructive to make a fast review on traversable wormholes.

\section{Wormhole Solutions}
Take into account a static and spherically symmetrical metric for wormhole spacetime geometry in the following manner:
\begin{equation}
ds^2=-e^{2\Phi(r)}dt^2+\frac{dr^2}{1-\frac{b(r)}{r}}+r^2(d\theta^2+\sin^2\theta d\phi^2), \label{13}
\end{equation}
where $r$ is the radial coordinate which runs in the range $r_0\leq r<\infty$, $0\leq\theta\leq\pi$ and $0\leq\phi\leq2\pi$ are the angular coordinates. $\Phi(r)$ and $b(r)$ denote the redshift function and shape function, respectively. To describe a traversable wormhole spacetime configuration in the framework of GR, generally speaking, four fundamental ingredients are needed \cite{8}:

$\star$ The NEC must be violated.

$\star$ Satisfy the so-called flaring-out conditions, namely, $b(r_0)=r_0$, $b'(r_0)<1$ and $b(r)<r$ when $r>r_0$.

$\star$ $\Phi(r)$ should be finite anywhere, in order that an horizon can be avoided.

$\star$ The asymptotically flatness should be satisfied, namely, $b/r\rightarrow0$ and $\Phi\rightarrow0$ when $r\rightarrow\infty$.

Utilizing the Einstein field equation, $G_{\mu\nu}=T_{\mu\nu}$, in an orthonormal reference frame, one can obtain the stress energy scenario as follows:
\begin{equation}
\rho=\frac{b'}{r^2}, \label{14}
\end{equation}
\begin{equation}
p_r=2\frac{\Phi'}{r}(1-\frac{b}{r})-\frac{b}{r^3}, \label{15}
\end{equation}
\begin{equation}
p_t=(1-\frac{b}{r})[\Phi''+(\Phi')^2-\Phi'\frac{b'r-b}{2r^2(1-b/r)}-\frac{b'r-b}{2r^3(1-b/r)}+\frac{\Phi'}{r}], \label{16}
\end{equation}
where $\rho(r)$ denotes the matter energy density, $p_r(r)$ the radial pressure, $p_t(r)$ the tangential pressure and the prime the derivative with respect to $r$. Utilizing the stress energy conservation equation, $T^{\mu\nu}_{\hspace{3mm};\nu} = 0$, we can obtain
\begin{equation}
p'_r=\frac{2}{r}(p_t-p_r)-\Phi'(\rho+p_r), \label{17}
\end{equation}
which can be thought of as the hydrostatic equation of equilibrium for the matter supporting a wormhole spacetime configuration or the relativistic Euler equation. In what follows, we will use the average dark energy equation of state $p=\omega_{av}\rho$ to calculate the wormhole solutions. Note that we have assumed that the value of $\omega_{av}(z_1,z_2)$ is a constant in the redshift range $[z_1,z_2]$ when implementing the constraint, therefore, the solutions we obtain will also lie in the redshift range $[z_1,z_2]$. To be more precise, $\omega_{av}=-1.5$ and $[z_1,z_2]=[0.32,2.34]$ is adopted throughout the calculating process, since $\omega_{av}=-1.5$ lies in the confidence levels of three different constraint results (see the horizontal green line in Fig. \ref{f1}).

\subsection{The Constant Redshift Function}
Consider the case of a constant redshift function $\Phi=C$, substituting the average dark energy equation of state $p=\omega_{av}\rho$ and Eq. (\ref{14}) into Eq. (\ref{15}), we obtain the following shape function
\begin{equation}
b(r)=r_0(\frac{r_0}{r})^{\frac{1}{\omega_{av}}}. \label{18}
\end{equation}
It is easy to verify $b(r)<r$ when $r>r_0$, which satisfies the usual flaring-out conditions. In what follows, evaluating at the wormhole throat $r_0$, we obtain
\begin{equation}
b'(r_0)=-\frac{1}{\omega_{av}}. \label{19}
\end{equation}
Furthermore, using the same value $\omega_{av}=-1.5$ based on the BAO observations in three different cases, one can easily get $b'(r_0)=0.67$. Since the corresponding redshift function $\Phi$ is finite everywhere and $b/r\rightarrow0$ when $r\rightarrow\infty$, this wormhole solution is both traversable and asymptotically flat. Hence, in principle, the dimension of the wormhole configuration can be very large.

As described and done in our previous works \cite{4,5,6}, we would like to continue investigating the intriguing topic, i.e., the traversability of the wormhole. In general, for a traveler in the spaceship who intends to journey through the wormhole throat, three necessary conditions are needed:

$\star$ The acceleration felt by the travelers should not exceed 1 Earth's gravitational acceleration $g_\star$:
\begin{equation}
\left|(1-\frac{b}{r})^{\frac{1}{2}}(\delta e^\Phi)'e^{-\Phi}\right|\leq g_\star.
\end{equation}

$\star$ The tidal acceleration should not exceed 1 Earth's gravitational acceleration $g_\star$:
\begin{equation}
\left|\eta^1\right|\left|(1-\frac{b}{r})[\Phi''+(\Phi')^2-\frac{b'r-b}{2r(r-b)}\Phi']\right|\leq g_\star,
\end{equation}
\begin{equation}
\left|\eta^2\right|\left|\frac{\delta^2}{2r^2}[v^2(b'-\frac{b}{r})+2(r-b)\Phi']\right|\leq g_\star,
\end{equation}
where $v$ is the traveler's velocity and  $\left|\eta^i\right|\approx2 m$ ($i=1,2$) is the distance between two arbitrary parts of the traveler's body (the size of the traveler) \cite{8}.

$\star$ The traverse time measured by the travelers and the observers who keep at rest at the space station should satisfy the several quantitative relationships:
\begin{equation}
\Delta\tau=\int^{+l_2}_{-l_1}\frac{dl}{v\delta},
\end{equation}
\begin{equation}
\Delta t=\int^{+l_2}_{-l_1}\frac{dl}{ve^\Phi},
\end{equation}
where $dl=(1-\frac{b}{r})^{-1/2}dr$ is the proper radial distance, and we consider that the space stations are located at a radius $r=a$, at $-l_1$ and $l_2$, respectively.

\begin{figure}
\centering
\includegraphics[scale=0.5]{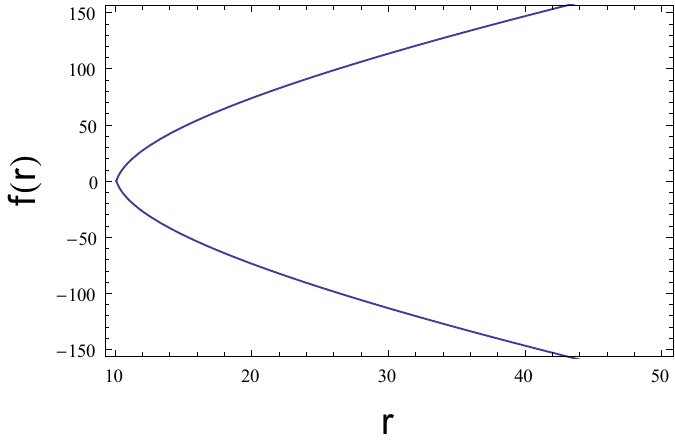}
\caption{The profile curve of the traversable wormhole.}\label{f2}
\end{figure}

\begin{figure}
\centering
\includegraphics[scale=0.5]{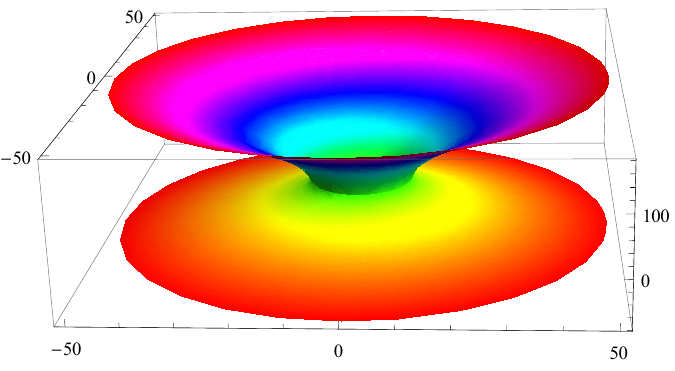}
\caption{The embedding diagram generated by rotating the profile curve about the vertical axis.}\label{f3}
\end{figure}

It is obvious that inequalities (20) and (21) is well satisfied in the case of a constant redshift function. Substituting Eqs. (18-19) into inequality (22) evaluated at the throat, neglect the substantial small term that contains the redshift $z$, considering a constant non-relativistic traversal velocity, namely, $\gamma\approx1$, we obtain the new expression for the velocity:

\begin{equation}
v\leqslant r_0\sqrt{\frac{\omega_{av} g_\star}{(\omega_{av}+1)}}, \label{20}
\end{equation}
where $v$ denotes the traversal velocity, $r_0$ the throat radius of the wormhole and $g_\star$ 1 Earth's gravitational acceleration. Note that we have chosen the height of a traverser as 2 m in this situation. In what follows, assuming $r_0=100$ m, we can easily obtain the traverse velocity $v\approx542.218$ m/s.
Furthermore, we can also easily have the traverse time $\Delta\tau\approx\Delta t\approx2R/v\thickapprox36.886s$ by assuming the junction radius $R=10000$ m.
It is worth noting that, from Eq. (\ref{20}), the traverse time for a traverser depends entirely on the wormhole spacetime geometry and cosmological background effects, i.e., the throat radius of the wormhole $r_0$, the earth gravitational field $g_\star$ and the average dark energy equation of state $\omega_{av}$. In addition, we think that it is necessary to preform the spacetime structure of the wormhole configuration. One can get the wormhole spacetime configuration through the function $f(r)$, which depicts the embedded surface of the wormhole. In general, this function is defined in the following manner
\begin{equation}
\frac{df}{dr}=\pm\frac{1}{\sqrt{\frac{r}{b(r)}-1}}. \label{21}
\end{equation}
In what follows, we can make plots for the profile function $f(r)$ and wormhole structure through solving this differential equation numerically (see Figs. (\ref{f2}-\ref{f3})). Notice that here we have assumed the radius of the wormhole throat $r_0=10$ m and the average dark energy equation of state $\omega_{av}=-1.5$.

\subsection{The Shape Function $b(r)=r_0+\frac{1}{\omega_{av}}(r_0-r)$}
Consider a special shape function $b(r)=r_0+\frac{1}{\omega_{av}}(r_0-r)$ and use the average dark energy equation of state $p=\omega_{av}\rho$, we obtain $\Phi'(r)=-\frac{1}{2r}$. Subsequently, integrating over $r$ on both sides, we get $\Phi(r)=-\frac{1}{2}\ln r+C$, where $C$ is an arbitrary integration constant. It is easy to find that this wormhole solution is non-asymptotically flat, since it diverges directly when $r\rightarrow\infty$. Hence, this wormhole solution is not traversable for a traveler. However, we can construct theoretically a traversable wormhole through matching an exterior flat vacuum geometry into the interior spacetime geometry supporting by phantom energy at a junction radius $r_j$. Hence, the constant $C$ is expressed as $C=\Phi(r_j)+\frac{1}{2}\ln(\frac{r_j}{r})$.

In what follows, we shall calculate out the amounts of the exotic matter threading the wormhole by utilizing the method of `` volume integral quantifier '' (VIQ) \cite{9}. For simplicity, one can obtain the amounts of the exotic matter by computing the definite integral $\int T_{\mu\nu}k^\mu k^\nu dV$, and here we just consider the finite range of the traversable wormhole $r\in[r_0, r_j]$. Subsequently, using the corresponding physical quantity $I=\int[p_r(r)+\rho]dV$, we have
\begin{equation}
I_V=\int^{r_j}_{r_0}(r-b)[\ln(\frac{re^{2\Phi}}{r-b})]'dr. \label{22}
\end{equation}
It follows that $I_V=(1+\frac{1}{\omega_{av}})(r_0-r_j)$. As before, continuing taking the average dark energy equation of state $\omega_{av}=-1.5$ from BAO data, we can have
\begin{equation}
I_V=0.33(r_0-r_j). \label{23}
\end{equation}
It is easy to demonstrate that the physical quantity $I_V\rightarrow0$ when $r_j\rightarrow r_0$. The result suggests that, in theory, one can construct a traversable wormhole with infinitesimal amounts of averaged-NEC (ANEC) violating dark energy fluid in the present scenario. Note that we have obtained the same result directly starting from BAO observations as previous works \cite{4,5,6,10}, which are based on concrete cosmological models.

\section{Stability Analysis}
In this section, for simplicity, we just perform the stability analysis of the first solution corresponding to the case of the constant redshift function. Using the usual Darmois-Israel formalism \cite{11,12}, the Lanczos equations can be expressed as
\begin{equation}
S^i_j=-\frac{1}{8\pi}(\kappa^i_j-\delta^i_j\kappa^k_k), \label{24}
\end{equation}
where $S^i_j$ and $\kappa^i_j$ denote, respectively, the surface stress energy tensor at the junction surface $\Sigma$ and the discontinuity of the extrinsic curvatures across the surface $\Sigma$. More detailed descriptions can be found in \cite{13}.

Considering the wormhole spacetime geometry Eq. (\ref{13}) and the Schwarzschild solution, we can express the nonzero components of the extrinsic curvatures as follows
\begin{equation}
K^{\tau+}_\tau=\frac{\frac{M}{a^2}+\ddot{a}}{\sqrt{1-\frac{2M}{a}+\dot{a}^2}}, \label{25}
\end{equation}
\begin{equation}
K^{\tau-}_\tau=\frac{\Phi'(1-\frac{b}{a}+\dot{a}^2)+\ddot{a}-\frac{\dot{a}^2(b-b'a)}{2a(a-b)}}{\sqrt{1-\frac{b(a)}{a}+\dot{a}^2}},  \nonumber
\end{equation}
\begin{equation}
K^{\theta+}_\theta=\frac{1}{a}\sqrt{1-\frac{2M}{a}+\dot{a}^2},  \nonumber
\end{equation}
\begin{equation}
K^{\theta-}_\theta=\frac{1}{a}\sqrt{1-\frac{b(a)}{a}+\dot{a}^2},  \nonumber
\end{equation}
where $a$ denotes the junction radius situated outside the event horizon ($a>2M$). Subsequently, according to the Lanczos equation, the surface energy density and the tangential surface pressure can be, respectively, expressed as
\begin{equation}
\sigma=-\frac{1}{4\pi a}(\sqrt{1-\frac{2M}{a}+\dot{a}^2}-\sqrt{1-\frac{b(a)}{a}+\dot{a}^2}), \label{26}
\end{equation}
and
\begin{equation}
P=\frac{1}{8\pi a}[\frac{1-\frac{M}{a}+\dot{a}^2+a\ddot{a}}{\sqrt{1-\frac{2M}{a}+\dot{a}^2}}-\frac{(1+a\Phi')(1-\frac{b}{a}+\dot{a}^2)+a\ddot{a}-\frac{\dot{a}^2(b-b'a)}{2(a-b)}}{\sqrt{1-\frac{b(a)}{a}+\dot{a}^2}}]. \label{27}
\end{equation}

Using the second contracted Gaussian-Kodazzi equation (or the `` ADM '' constraint) and Lanczos equations, we can have
\begin{equation}
\sigma'=-\frac{2}{a}(\sigma+P)+\Xi, \label{28}
\end{equation}
where $\Xi$ is provided by
\begin{equation}
\Xi=-\frac{1}{4\pi a^2}[\frac{b'a-b}{2a(1-\frac{b}{a})+a\Phi'}]\sqrt{1-\frac{b}{a}+\dot{a}^2}. \label{29}
\end{equation}
It is worth noting that Eq. (\ref{28}), namely, the radial derivative of the surface energy density, plays an important role in characterizing the stability regions of the wormhole solutions.

One can also obtain the thin shell's equation of motion after rearranging Eq. (\ref{26}) as follows
\begin{equation}
\dot{a}^2+V(a)=0, \label{30}
\end{equation}
with the potential provided by
\begin{equation}
V(a)=1+\frac{2Mb(a)}{m_s^2}-[\frac{m_s}{2a}+\frac{M+\frac{b(a)}{2}}{m_s}]^2, \label{31}
\end{equation}
where $m_s=4\pi a^2$ denotes the surface mass of the thin shell. For conveniences of calculations, we define two new functions as follows
\begin{equation}
F(a)=1-\frac{\frac{b(a)}{2}+M}{a}, \label{32}
\end{equation}
\begin{equation}
G(a)=\frac{M-\frac{b(a)}{2}}{a}, \label{33}
\end{equation}
in order that the potential can be rewritten as
\begin{equation}
V(a)=F(a)-(\frac{m_s}{2a})^2-(\frac{aG}{m_s})^2. \label{34}
\end{equation}
Following \cite{13,14}, we get the usual equilibrium relation by linearizing around a stable solution situated at $a=a_0$
\begin{equation}
\Gamma=(\frac{a_0}{m_s})[F'-2(\frac{a_0G}{m_s})(\frac{a_0G}{m_s})'], \label{35}
\end{equation}
and the stability condition of the solution
\begin{equation}
\Psi-\Gamma^2>(\frac{m_s}{2a})(\frac{m_s}{2a})'', \label{36}
\end{equation}
where $\Psi$ can be expressed as
\begin{equation}
\Psi=\frac{F''}{2}-[(\frac{aG}{m_s})']^2-(\frac{aG}{m_s})(\frac{aG}{m_s})''. \label{37}
\end{equation}
The former can be used in determining the master equation and the latter is obtained by using the condition $V''(a_0)>0$. Furthermore, utilizing the definition $m_s=4\pi a^2\sigma$, the radial derivative of $\sigma'$ and Eq. (\ref{21}), one can express the master equation as
\begin{equation}
4\pi\sigma'\lambda=\Upsilon-(\frac{m_s}{2a})'', \label{38}
\end{equation}
where $\lambda$ is defined as $\lambda=P'/\sigma'$ and $\Upsilon$ is given by
\begin{equation}
\Upsilon=2\pi a\Xi'+\frac{4\pi(\sigma+P)}{a}. \label{39}
\end{equation}
In the following context, we would like to use $\lambda$ as a parameterization of the stable equilibrium configurations in order to avoid specifying a surface equation of state. Subsequently, replacing Eq. (\ref{38}) in Eq. (\ref{36}), one can rewrite the master equation as
\begin{equation}
\lambda_0\sigma'm_s>\Theta, \label{40}
\end{equation}
where $\lambda_0=\lambda(a_0)$ and $\Theta$ is defined as
\begin{equation}
\Theta=\frac{a_0(\Gamma^2-\Psi)}{2\pi}+\frac{m_s\Upsilon}{4\pi}. \label{41}
\end{equation}
Therefore, it is clear that the stability regions of the wormhole solutions can be well described by the following conditions
\begin{equation}
\lambda_0>\tilde{\Theta}, \quad if \quad \sigma'm_s>0, \label{42}
\end{equation}
\begin{equation}
\lambda_0<\tilde{\Theta}, \quad if \quad \sigma'm_s<0, \label{43}
\end{equation}
with the definition $\tilde{\Theta}\equiv\frac{\Theta}{\sigma'm_s}$.

Now, we can apply this method into the case $\Phi=C$. To be more precise, the factor $\Xi$ which is related to the net discontinuity of the momentum flux impinging on the thin shell, is expressed as
\begin{equation}
\Xi=\frac{1}{8\pi a_0^2}\frac{(\frac{1+\omega_{av}}{\omega_{av}})(\frac{r_0}{a_0})^{\frac{1+\omega_{av}}{\omega_{av}}}}{\sqrt{1-(\frac{r_0}{a_0})^{\frac{1+\omega_{av}}{\omega_{av}}}}}. \label{44}
\end{equation}
The factor related to the equilibrium relation is written as
\begin{equation}
\Gamma=\frac{1}{2a_0}[\frac{1}{2}\frac{(\frac{1+\omega_{av}}{\omega_{av}})(\frac{r_0}{a_0})^{\frac{1+\omega_{av}}{\omega_{av}}}}{\sqrt{1-(\frac{r_0}{a_0})^{\frac{1+\omega_{av}}{\omega_{av}}}}}-\frac{\frac{M}{a_0}}{\sqrt{1-\frac{2M}{a_0}}}], \label{45}
\end{equation}
In the meanwhile, the radial derivative of the surface energy density $\sigma'(a_0)$ can be expressed as
\begin{equation}
\sigma'=\frac{1}{4\pi a_0^2}(\frac{1-\frac{3M}{a_0}}{\sqrt{1-\frac{2M}{a_0}}}-\frac{1-(\frac{1+\omega_{av}}{2\omega_{av}})(\frac{r_0}{a_0})^{\frac{1+\omega_{av}}{\omega_{av}}}}{\sqrt{1-(\frac{r_0}{a_0})^{\frac{1+\omega_{av}}{\omega_{av}}}}}). \label{46}
\end{equation}
Since the explicit forms of the remaining functions $\Upsilon$, $\Psi$ and $\Theta$ are substantially cumbersome, we would not like to write them analytically. Nonetheless, it is very constructive and necessary to exhibit the stability regions graphically.

In the following context, we will separate the cases of $b(a_0)<2M$ and $b(a_0)>2M$, which correspond to $m_s>0$ and $m_s<0$, in order to perform the stability analysis of the wormhole solution. As noted in \cite{13,14}, we would like to relax the condition that the surface energy density is positive. It is noteworthy that we are dealing with the exotic matter, and for $\sigma<0$, the weak energy condition (WEC) is directly violated.

\begin{figure}
\centering
\includegraphics[scale=0.5]{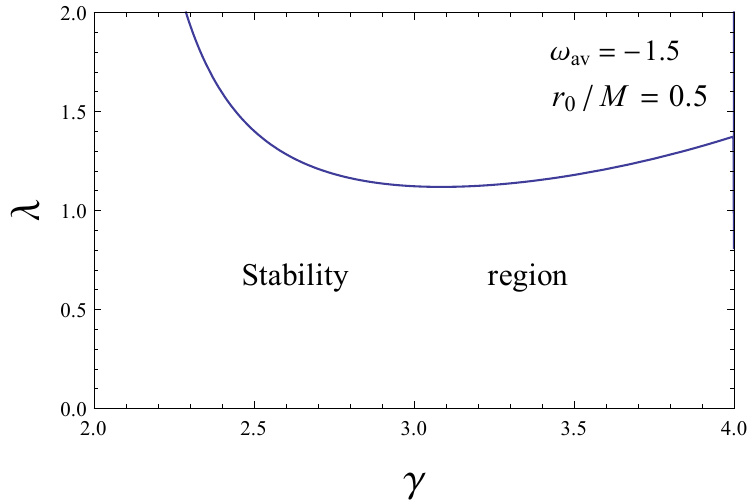}
\includegraphics[scale=0.5]{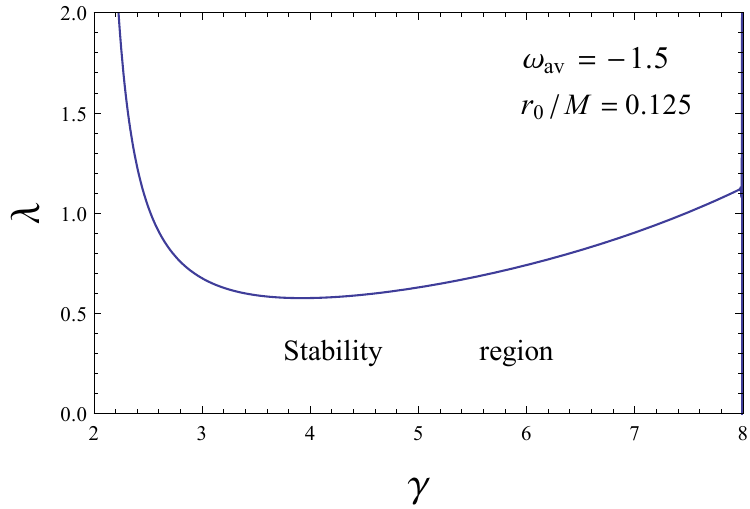}
\caption{Plots for a positive surface energy density, i.e., $b(a_0)<2M$. We consider the case $\omega_{av}=-1.5$ and define $\gamma=a_0/M$ for both cases. The left and right panels represent $r_0/M=0.5$ and $r_0/M=0.125$, respectively. The stability regions are exhibited beneath the solid curves.}\label{f6}
\end{figure}

For the case of $b(a_0)<2M$, in general, we need to impose a condition $r_0<2M$, in order that the matching radius lies outside the event horizon $a_0>2M$. Subsequently, we can naturally have
\begin{equation}
2M<a_0<2M(\frac{2M}{r_0})^{-(1+\omega_{av})}. \label{47}
\end{equation}
By fixing the value of the average dark energy equation of state $\omega_{av}=-1.5$, we take into account the following cases: $r_0/M=0.5$, in order that $2<a_0/M<4$; for $r_0/M=0.125$, we get $2<a_0/M<8$. Furthermore, we exhibit the respective stability regions in Fig. \ref{f6}. From Eqs. (\ref{43}) and (\ref{46}), we have $\sigma'<0$ and $m_s\sigma'<0$. Hence, the stability regions lie below the solid curves in the plots of Fig. \ref{f6}. It is worth noticing that the stability regions increase for increasing values of $r_0/M$, and for decreasing values of $r_0/M$, regardless of the increasing range $a_0$, the values of the parameter $\lambda_0$ are further restricted.

\begin{figure}
\centering
\includegraphics[scale=0.5]{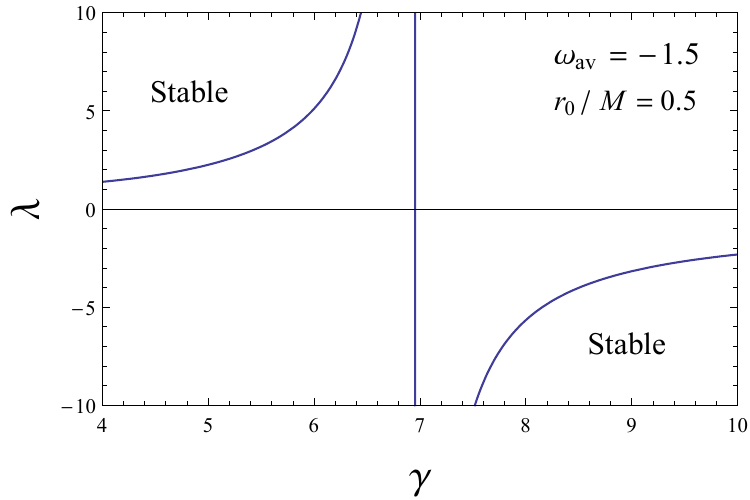}
\includegraphics[scale=0.5]{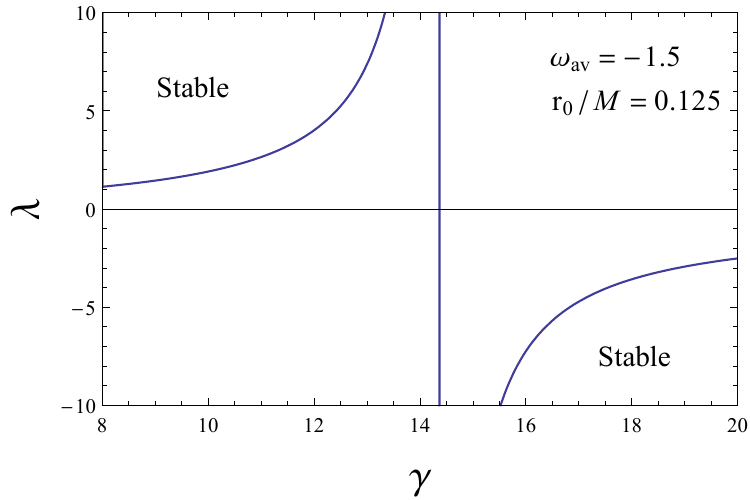}
\caption{Plots for a negative surface energy density, i.e., $b(a_0)>2M$. We consider the case $\omega_{av}=-1.5$ and define $\gamma=a_0/M$ for both cases. The left and right panels represent $r_0/M=0.5$ and $r_0/M=0.125$, respectively. The stability regions are exhibited above the first solid curve and beneath the second solid curve.}\label{f7}
\end{figure}

For the case of $b(a_0)<2M$, the surface mass of the thin shell $m_s(a_0)<0$. Here we will discuss the cases of $r_0<2M$ and $r_0>2M$ separately.

For $r_0<2M$, the range of the junction radius can be expressed as
\begin{equation}
a_0>2M(\frac{2M}{r_0})^{-(1+\omega_{av})}. \label{48}
\end{equation}
By implementing some numerical calculations, we find that $\sigma'$ has a real positive root $R$ in the range of Eq. (\ref{48}), i.e., $\sigma'|_R=0$. As described in \cite{13,14}, we also demonstrate that $\sigma'<0$ for $2M(\frac{2M}{r_0})^{-(1+\omega_{av})}<a_0<R$, and $\sigma'>0$ for $a_0>R$. Therefore, the stability regions can be expressed as
\begin{equation}
\lambda_0>\tilde{\Theta}, \quad if \quad 2M(\frac{2M}{r_0})^{-(1+\omega_{av})}<a_0<R, \label{49}
\end{equation}
\begin{equation}
\lambda_0<\tilde{\Theta}, \quad if \quad a_0>R. \label{50}
\end{equation}

Similarly, fixing $\omega_{av}=-1.5$, we also consider the specific cases: $r_0/M=0.5$, in order that $a_0/M>4$; for $r_0/M=0.125$, we get $a_0/M>8$. At the same time, the asymptotes $\sigma'|_R=0$ for these two cases exist at $R/M\simeq6.95$ and $R/M\simeq14.37$, respectively (see Fig. \ref{f7}). It is worth noticing that the range of $a_0$ decreases for increasing values of $r_0/M$, and the values of the parameter $\lambda_0$ are less restricted. Furthermore, it seems that there is no apparent increasing tendency for the stability regions with increasing values of $r_0/M$.

For $r_0>2M$, one can easily have $\sigma'>0$ and $m_s\sigma'<0$. Subsequently, according to Eq. (\ref{43}), we find that the values of the parameter $\lambda_0$ are always negative. In addition, with increasing values of $r_0/M$, the range of $a_0$ decreases and the values of the parameter $\lambda_0$ is also less restricted.
\section{Discussions and conclusions}
Since the elegant discovery that our universe is undergoing the phase of accelerated expansion, recently, theorists have gradually paid more and more attention to the exotic spacetime configurations, especially, the renewed field---wormholes.

In our previous works \cite{4,5,6}, we are dedicated to interpret that dark energy may act as the matter source of the wormhole spacetime configurations. Additionally, we have explored the related properties of the wormhole solutions and the energy conditions for a given cosmological model in a point view of observational cosmology. However, the previous investigations all depend on the concrete cosmological models, and may do not have enough representativeness and persuasiveness. Thus, in this study, we are aiming at investigating the model-independent traversable dark energy wormholes from baryon acoustic oscillations.

At first, we preform the model-independent method and the statistical constraints on the average dark energy equation of state by only using the BAO data. We find that our constraining results are different from those in paper \cite{2}. We think that the discrepancies can be attributed to the different choices of the flat prior for the parameter pair in the above-mentioned three cases. In the meanwhile, the discrepancy will also not affect the correctness of our conclusions about the wormholes. Subsequently, two specific wormhole solutions directly starting from BAO data are obtained. For the case of constant redshift function, we analyze the traversabilities of the corresponding wormhole spacetime configuration, and find that the traverse time for a traverser depends entirely on the wormhole spacetime geometry and cosmological background effects, i.e., the throat radius of the wormhole $r_0$, the earth gravitational field $g_\star$ and the average dark energy equation of state $\omega_{av}$. In addition, to characterize the wormhole spacetime structure better, we also exhibit the corresponding embedded diagram for the wormhole configuration. For the second case, we use the VIQ method to work out the amounts of the exotic matter threading the wormhole and find that one can construct theoretically a traversable wormhole with infinitesimal amounts of ANEC violating phantom fluid in the present scenario. At last, as a concrete performance, we exhibit the stability analysis for the first case in the absence of the thin shell, and find that the stable equilibrium configurations may increase for increasing values of the throat radius of the wormhole in the cases of a positive and a negative surface energy density. Furthermore, we also find that in the case of $b(a_0)<2M$, for decreasing values of the throat radius of the wormhole, regardless of the increasing range $a_0$, the values of the parameter $\lambda_0$ are further restricted. However, in the case of $b(a_0)>2M$, the values of the parameter $\lambda_0$ are less restricted for decreasing values of the throat radius of the wormhole. It is worth noting that the square root of the parameter $\lambda$ is usually interpreted as the speed of sound, in order that one could expect $0<\lambda\leqslant1$ which corresponds to the requirement that the speed of sound should not exceed the speed of light. Nonetheless, in the presence of the exotic matter, this requirement can be relaxed since we still have little knowledge about the nature of the dark energy.

Note that the obtained wormhole solutions are static and spherically symmetrical metric and we assume $\omega_{av}$ to be a constant between different redshifts when placing constraints, therefore, these wormhole solutions can be interpreted as stable and static phantom wormholes configurations at some certain redshift which lies in the range [0.32, 2.34]. Furthermore, these wormhole solutions are model-independent. This means that one can directly extract useful astrophysical information from observational data, or conversely, one can effectively probe the cosmological background evolution by astrophysical scale observations.

In the future, we expect and believe that more and more highly accurate observations can bring us evolutionary cognition about various celestial bodies, for instance, black holes, wormholes, white dwarfs, pulsars, quasars, etc.

\section{acknowledgements}
The author warmly thanks the anonymous referee for helping improve the manuscript and also thanks Professors Bharat Ratra and Sergei. D. Odintsov for helpful feedbacks on cosmology and astrophysics.

\end{document}